\documentclass[times, twoside, watermark]{zHenriquesLab-StyleBioRxiv}
\usepackage{blindtext}

\leadauthor{K. M. Kabusure, P. Piskunen}

\title{Raman Enhancement in Bowtie-Shaped Aperture-Particle Hybrid Nanostructures Fabricated with DNA-Assisted Lithography}
\shorttitle{Preprint}

\author[1,$\dagger$]{Kabusure M. Kabusure}
\affil[1]{Department of Physics and Mathematics, University of Eastern Finland, Yliopistokatu 2, P.O. Box 111, 80101, Joensuu, Finland}

\author[2,$\dagger$]{Petteri Piskunen}
\affil[2]{Biohybrid Materials, Department of Bioproducts and Biosystems, Aalto University, P.O. Box 16100, 00076, Aalto, Finland}

\author[1]{Jiaqi Yang}

\author[2,3,4,\Letter]{Veikko Linko}
\affil[3]{LIBER Center of Excellence, Aalto University, P.O. Box 16100, 00076, Aalto, Finland}
\affil[4]{Institute of Technology, University of Tartu, Nooruse 1, 50411, Tartu, Estonia}

\author[1,\Letter]{Tommi K. Hakala}

\affil[$\dagger$]{These authors contributed equally}

\begin{document}

\maketitle

\begin{abstract}
We report on efficient surface-enhanced Raman spectroscopy (SERS) supporting substrates, which are based on DNA-assisted lithography (DALI) and a layered configuration of materials. In detail, we used nanoscopic DNA origami bowtie templates to form hybrid nanostructures consisting of aligned silver bowtie-shaped particles and apertures of similar shape in a silver film. We hypothesized that this particular geometry could facilitate a four-fold advantage in Raman enhancement compared to common particle-based SERS substrates, and further, we verified these hypotheses experimentally and by finite difference time domain simulations. In summary, our DALI-fabricated hybrid structures suppress the background emission, allow emission predominantly from the areas of high field enhancement, and support additional resonances associated with the nanoscopic apertures. Finally, these nanoapertures also enhance the fields associated with the resonances of the underlying bowtie particles. The versatility and parallel nature of our DNA origami-based nanofabrication scheme and all of the above-mentioned features of the hybrid structures therefore make our optically resonant substrates attractive for various SERS-based applications.
\end {abstract}


\begin{corrauthor}
veikko.pentti.linko@ut.ee /
tommi.hakala@uef.fi

\end{corrauthor}

\section*{Introduction}

Various metallic nanostructures have been intensively studied owing to their ability to locally increase the incoming electromagnetic field intensity \textit{via} plasmon resonances.\cite{Tabor2009plasmonic,novotny2011antennas} Single metal nanoparticles,\cite{hao2004electromagnetic} metal particle arrangements with nanoscale gaps between the objects,\cite{hao2004electromagnetic,rechberger2003optical,sundaramurthy2005field} as well as apertures in metal films,\cite{Ebbesen1998,martin2001theory,Djaker2010surface} have all been shown to exhibit optically intriguing properties exploitable in applications such as sensing.\cite{mejia2018plasmonic} From these examples, particularly structures with nanoscale gaps, such as bowtie antennas\cite{sundaramurthy2005field,Fromm2004gap,kinkhabwala2009large} exhibiting intense plasmonic hotspots,\cite{Dodson2013optimizing} are attractive for surface-enhanced Raman spectroscopy (SERS)\cite{kneipp2002surface,hering2008sers} as the Raman enhancement factor scales with the fourth power of the electric field enhancement.

However, for the Raman enhancement, also plasmonic apertures, \textit{i.e}.\ metallic films perforated with nanoscopic holes,\cite{Ebbesen1998,kim1999control,martin2001theory,Dintinger2006molecule,Wei2008individual,Guo2008optical,Djaker2010surface,Schnell2010phase,Jensen2016optical,Chen2017bridged} may become highly attractive options. The reasoning is that the metal layer could potentially filter and suppress the background signal of the Raman measurement, consequently allowing the light to emanate only from the regions of high field enhancement. This could be very beneficial, as the Raman signal of interest may easily get obscured by the high background emission intensity.\cite{matousek2002fluorescence,wei2015review} 

Conventionally, metallic nanostructures have been fabricated employing top-down approaches. Recently, however, affordable and highly parallel bottom-up based methods have become increasingly sophisticated.
As a prime example, utilizing self-assembled DNA templates have allowed fabrication of optically active materials by precision-positioning of nano\-particles\cite{kuzyk2012dna, acuna2012fluorescence,thacker2014dna,roller2017hot,pilo-pais2017sculpting,Heck2017gold,zhan2018dna,shen2018dna,kuzyk2018nanophotonics,yesilyurt2021emission,tapio2021versatile} or by transferring the spatial information of the DNA template to entirely inorganic structures.\cite{Gates2014dna-templated,bathe2019roadmap,Heuer-Jungemann2021engineering,martynenko2021dna,hui2022dna}
Following these concepts, we have previously developed techniques that could take advantage of both the bottom-up -based DNA nanotechnology and the top-down approaches in fabrication of such optically resonant substrates. For instance, we have combined DNA origami nanostructures\cite{rothemund2006folding, douglas2009self, dey2021dna} as patterning templates with common micro-/nanofabrication schemes (such as thin film deposition and etching) to develop two techniques: DNA-assisted lithography (DALI)\cite{shen2018plasmonic} and the more versatile biotemplated lithography of inorganic nanostructures (BLIN).\cite{Piskunen2021Biotemplated} With these, we have previously patterned transparent surfaces with \textit{e.g}.\ bowtie-shaped metal nanoparticles with well-defined nanogaps (<10~nm) and demonstrated their feasibility in Raman enhancement.\cite{shen2018plasmonic,kabusure2022optical} Owing to their highly parallel and affordable fabrication processes, DALI and BLIN may, in general, serve as intriguing alternatives to the more conventional nanopatterning approaches. However, these methods do not support aperture fabrication which could be beneficial in developing even more efficient SERS substrates as discussed above.

 \begin{figure*}[t!]
    \centering
    \includegraphics[width=0.78\textwidth]{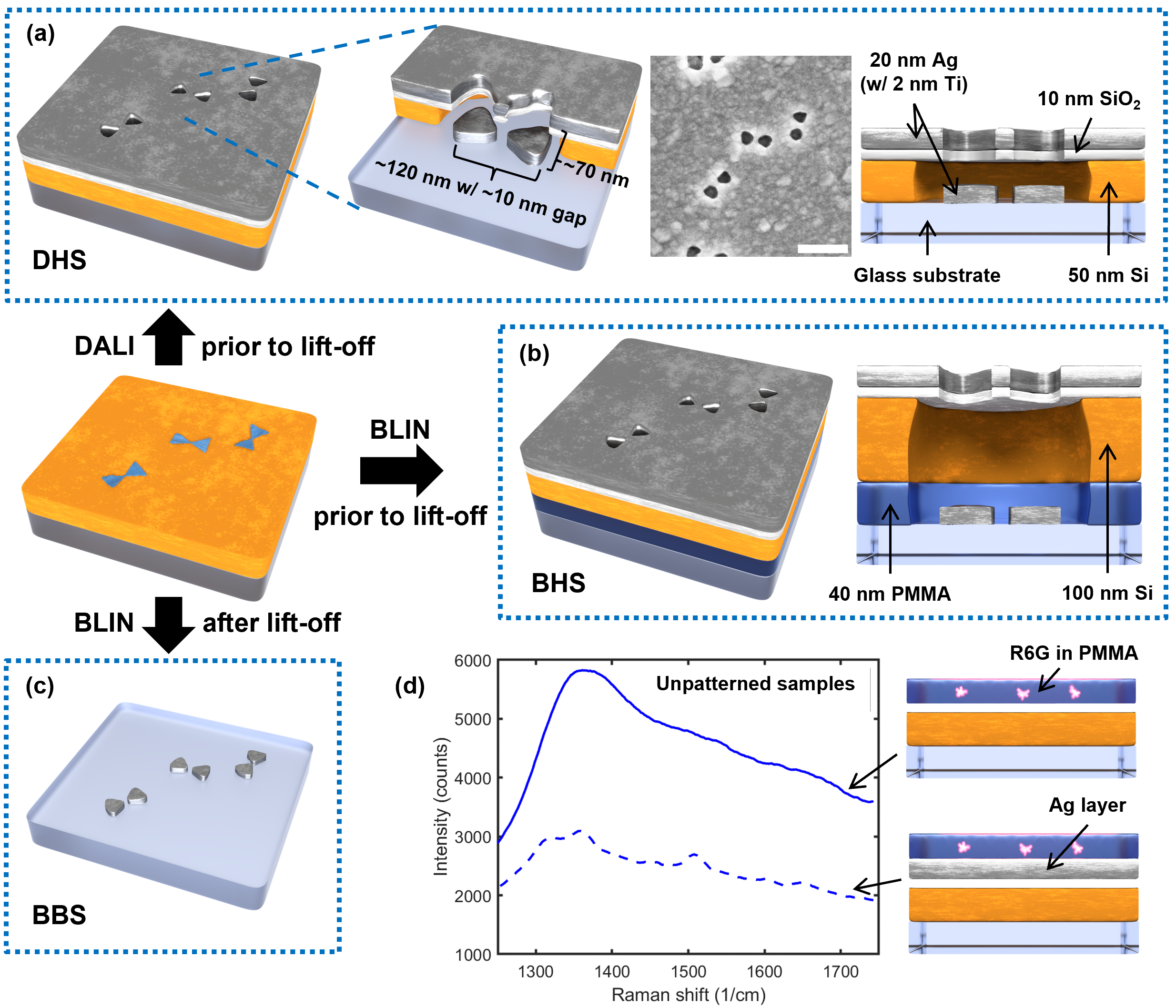}
    \caption{An overview of the optically resonant substrates used in this study. All fabrication processes start with a silicon-coated glass (or glass-PMMA) substrate on which the DNA origami bowties have been deposited (middle left). The arrows indicate the fabrication processes used to achieve the end products a, b, and c. (a) \textit{Left}: A DALI-fabricated hybrid structure (DHS) with a zoomed-in part showing the aperture-particle pair dimensions and positioning. \textit{Middle}: Scanning electron microscope (SEM) image of the sample; the scale bar is 200~nm. \textit{Right}: Cross-sectional model and dimensions of the bowtie aperture and particle features. (b) \textit{Left}: A BLIN-fabricated hybrid structure (BHS). \textit{Right}: Cross-sectional model and dimensions of the bowtie aperture and particle features. (c) BLIN-fabricated bowtie structures (BBS). (d) Comparison of the background intensities of the Raman signal for the unpatterned samples with and without a silver layer when coated with poly(methyl methacrylate) (PMMA) laced with rhodamine 6G (R6G) dye. These samples contain the same material layers and thicknesses as the DHS sample (subfigure a).}
    \label{scheme}
\end{figure*}

In this article, we show that we can modify the previous DNA-assisted lithography scheme in a way that results in a hybrid structure consisting of both aligned silver bowtie particles and nanoscale apertures of similar shape in a silver film (see DALI-fabricated hybrid structure, DHS, in \textbf{Figure~\ref{scheme}a}). We envision that this kind of a hybrid structure may exhibit very strong Raman enhancement that emerges from the intense plasmonic hotspots of the bowtie particles and the apertures as well as from spatial filtering properties of the aperture layer, allowing only the regions of high field enhancement to contribute to Raman signal.

In detail, we present four hypotheses and further show that our hybrid structure can significantly enhance Raman signals \textit{via} four separate mechanisms (four-fold advantage): \textit{Hypothesis 1 (H1):} The background emission can be suppressed by the aperture-containing metal film. \textit{Hypothesis 2 (H2):} The apertures allow light emission mainly from the areas of high field enhancement, a highly desirable feature for any practical implementation of Raman substrates. \textit{Hypothesis 3 (H3):} The apertures also support additional plasmonic effects that can result in significant field enhancements as such. \textit{Hypothesis 4 (H4):} The presence of nanoapertures can further enhance the fields associated with the resonances of the underlying bowtie particles.

To test and verify these hypotheses experimentally, we prepared several control samples for DHS and compared their performance in the detection of rhodamine 6G (R6G), a dye commonly employed in SERS experiments. We used BLIN processing to fabricate similar sandwich-like hybrid structures on glass but with a sacrificial layer included (BLIN-fabricated hybrid structures, BHS, see \textbf{Figure~\ref{scheme}b}). This also allows completion of a lift-off step to yield bare bowtie particles on the substrate (BLIN-fabricated bowtie structures, BBS, see \textbf{Figure~\ref{scheme}c}). In addition to these, we created unpatterned samples with the same layer composition as in DHS, both with and without the metal film, to study the role of the thin films in the reduction of background emission (\textbf{Figure~\ref{scheme}d}). Furthermore, to separate the contributions of the individual and combined effects emerged from the bowtie particles and the apertures, we performed detailed finite difference time domain (FDTD simulations) for the DHS samples.

\section*{Experimental Section}

\subsection*{Fabrication of Bowtie-Shaped Apertures and Particles}

The aperture-particle hybrid structures (DHS and BHS) and bowtie structures (BBS) were created using either BLIN\cite{Piskunen2021Biotemplated} or DALI techniques,\cite{shen2018plasmonic,Piskunen2019dna} and by employing DNA origami bowties as templates (designs, structural validation and folding protocols for the DNA origami bowties have been reported elsewhere.\cite{shen2018plasmonic}) In brief, the DHS structures were fabricated by adapting DALI on an ordinary glass substrate and omitting the final lift-off process (\textbf{Figure \ref{scheme}a}). To compare these DHS patterns to previously fabricated similar features,\cite{kabusure2022optical} BHS samples were fabricated with BLIN by also omitting the final lift-off step (\textbf{Figure~\ref{scheme}b}). Conversely, the BBS structures were created by performing the complete BLIN process on glass as shown earlier (\textbf{Figure~\ref{scheme}c}).\cite{kabusure2022optical} The full design, folding instructions and structural validation for the used DNA origami bowties are available in previous works.\cite{shen2018plasmonic,Piskunen2021Biotemplated} All materials and their sources are listed in the \textbf{Supporting Information Table~S1} and employed tools in \textbf{Table~S2}. The process parameters for all fabrication steps are given in \textbf{Table~S3}.

To begin processing of all samples, 0.5~mm thick borosilicate glass slides were first diced into$~10 \times 10$ mm chips. The chips were then cleaned by soaking in hot acetone (52~\textdegree C) for 1.5~h followed by an acetone rinse and 1~min sonication in room temperature acetone. After sonication, the chips were rinsed once more with acetone, then submerged in and rinsed with isopropanol (IPA) and, finally, immediately dried with a N$_2$ flow. Next, in the case of DHS, a 50~nm a-Si layer was deposited on the cleaned glass using plasma-enhanced chemical vapor deposition (PECVD). Meanwhile, to prepare the BHS and BBS samples, instead of immediate a-Si PECVD, the chips were first spincoated with 40~nm of sacrificial poly(methyl methacrylate) (PMMA), the PMMA was vacuum-cured, and finally, a 100~nm of a-Si was deposited on the PMMA film. O$_2$ plasma treatment was then performed with a reactive ion etching (RIE) tool on all sample types to generate negative surface charges on the deposited a-Si and thus enable attachment of DNA origami templates in the next fabrication step.

Then, a solution of DNA origami in Mg$^{2+}$ supplemented folding buffer (FOB) was prepared (5 nM bowtie DNA origami in $1\times$ TAE buffer (40~mM Tris, 19~mM acetic acid, 1~mM ethylenediaminetetraacetic acid (EDTA)) with 100 mM Mg$^{2+}$ at pH $\sim$8.3)) as shown earlier,\cite{Piskunen2021Biotemplated} and 10~$\mu$l of the solution was drop cast on the plasma-treated a-Si surfaces. The origami solution was left to incubate, covered, in ambient conditions for 5 min and then the surfaces were washed three times with 100~$\mu$l of ddH$_2$O. After washing, the chips were dried under a N$_2$ flow. The 5~nM DNA origami concentration was chosen to avoid overcrowding and collapse of the template\cite{Piskunen2021Biotemplated} and to enable easier comparison to previously fabricated bowtie particles.\cite{kabusure2022optical} The surface-attached templates were then used in the selective growth\cite{surwade2013nanoscale, zhou2015mechanistic,shen2018plasmonic,shen2018dna-assisted} of a SiO$_2$ mask layer as detailed.\cite{Piskunen2021Biotemplated} A $\sim$20-h growth time was chosen to overgrow the thin waist feature in the bowties and to thus form gapped bowtie shapes.

Next, RIE was used to pierce the SiO$_2$ and a-Si layers (as well as the PMMA film in the BHS and BBS samples) to expose the underlying glass substrate, followed by physical vapor deposition (PVD) of Ti (2~nm) and Ag (20~nm) in ultra-high vacuum. Unlike in the previous DALI\cite{shen2018plasmonic} and BLIN\cite{Piskunen2021Biotemplated} techniques, no lift-off was performed after metal deposition for the DHS and BHS chips, which resulted in a Ag film with gapped bowtie-shaped apertures and correspondingly shaped self-aligned particles on the initial substrate (see Figure 1a and b). The fabricated features were imaged with scanning electron microscopy (SEM) (\textbf{Figure~\ref{scheme}a} and \textbf{Supporting Information Figure~S1}). After fabrication and imaging, the chips were spincoated with a vacuum cured, $\sim$40~nm thick PMMA layer to inhibit further oxidation of the Ag film prior to measurements.

\subsection*{Unpatterned Control Samples}
Two unpatterned control samples shown in \textbf{Figure~\ref{scheme}d} were prepared to investigate the optical responses of the used film configuration and the filtering effect (background suppression) of the employed Ag film (\textbf{Figure~\ref{scheme}d}). The samples were fabricated by following the same protocol as in the DHS sample fabrication, but the DNA origami template attachment and etching steps were omitted from the process to yield an unpatterned, but otherwise identical, stack of materials. One of the control samples was coated with Ti and Ag, while the other one was left without the metal films. These control samples were also coated with a protective PMMA layer similarly as the other samples to help preserve them before Raman measurements.

\subsection*{Sample Preparation for Raman Experiments} 

The protecting PMMA layer was first removed by immersing the samples into acetone  followed by a sonication step for $\sim$10~min. Samples were then cleaned by isopropanol (IPA) for $\sim$5~min to remove the remained PMMA residues and blow dried with a N$_2$ flow. Then, 1~ml of rhodamine 6G (R6G) solution (5 mg of R6G powder dissolved in 1 ml of ethanol) was mixed with 3.5 ml of PMMA A3 (3 \% 950K PMMA in anisole (w/v) prepared by diluting 1~ml of PMMA A11 with 2.5 ml of anisole), and finally, the structures were spin-coated with this R6G solution using a spin coater at 3,000 rpm for 30 s to produce a PMMA layer thickness of $\sim$120~nm.\cite{kabusure2022optical}

\subsection*{Raman Measurement}

A commercial Renishaw Invia Reflex Raman microscope accompanied with WiRE~$\textsuperscript{TM}$ software was used to measure the Raman signals of R6G spin-coated on the structures. The sample was imagined using white light and a~50$\times$ objective lens. After selecting the area of interest to measure, the laser source was switched on to allow illumination of 785~nm laser excitation wavelength on the sample. The centre wavelength of 1500~cm$^{-1}$,  diffraction grating of 1200~l/mm, and 10 s of exposure time were set, while constantly controlling the laser power to achieve optimal conditions for taking the measurements. 
All Raman spectra were averaged from 16 measurements covering an area of 100 $\times$ 100~$\mu$m$^2$.

\begin{figure*}[t!]
    \centering
    \includegraphics[width=0.78\textwidth]{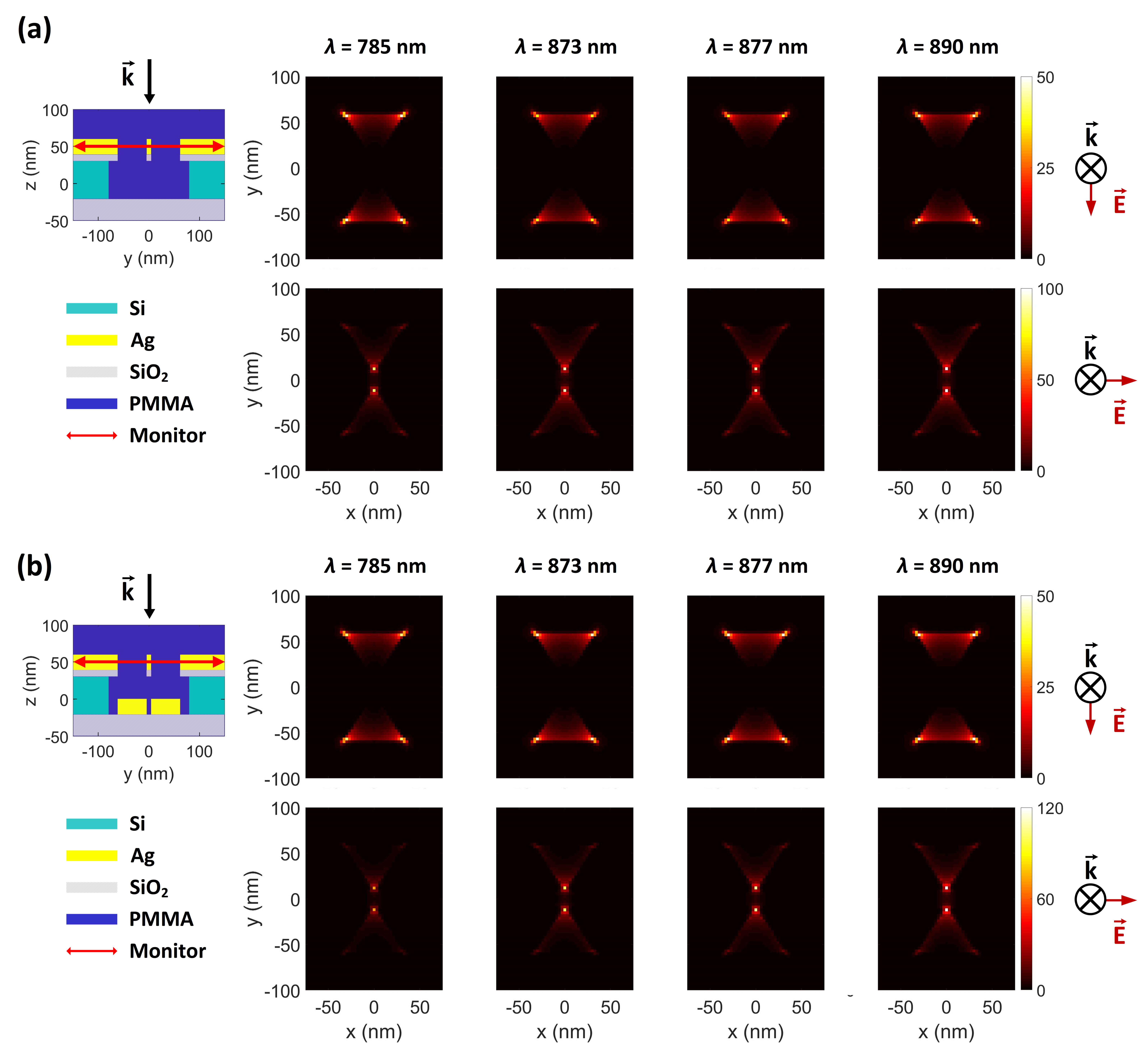}
    \caption{FDTD simulations showing the electric field intensities at the apertures. The monitored area is indicated in the inset. (a) Aperture only (i.e.\ DHS without a bowtie particle) as a control. (b) DHS sample.  The intensities are shown at the Raman excitation wavelength (785 nm) as well as at the Raman transition wavelengths of Rhodamine 6G (873 nm, 877 nm, and 890 nm) for both longitudinal (top panel) and transverse (bottom panel) polarizations.}
    \label{aperturesim}
\end{figure*}

\subsection*{Finite Difference Time Domain (FDTD) Simulations}

To investigate the electric field intensity enhancements (FEs) of the proposed DHS structure, we performed full-wave simulations using the finite-difference time domain (FDTD) technique in Lumerical simulation software (Ansys). We used stabilized perfectly matched layer (PML) as simulation boundaries to minimize reflections, which guaranteed better stability for the simulation and therefore more accurate results. The simulation waveband was chosen to be 100--1500~nm to cover potential Raman excitation wavelengths. The field intensity profiles were resolved for both transversely and longitudinally polarized light (x- and y-polarization, respectively; normal to the incident light). The intensity distributions are shown separately in \textbf{Figure~\ref{aperturesim}} and \textbf{Figure~\ref{bowtiesim}}, based on their different monitor locations (see also \textbf{Supporting Information Figure~S1}).

\section*{Results and Discussion}

As described in the introduction, we hypothesized in total four different mechanisms for Raman enhancement in our DHS-based system. To verify the first one, \textit{H1}, we compared the Raman signals from the unpatterned samples with and without the silver layer and showed that the layer indeed blocks the background emission from the substrate effectively (\textbf{Figure~\ref{scheme}d}). It is noteworthy that even though the silver layer expectedly enhances the Raman signal of the R6G and the characteristic peaks start to appear, the overall signal is significantly reduced due to the filtering effect by the silver layer. 

Further, to test the next hypotheses, we separated the individual effects of the apertures and bowtie particles by performing FDTD simulations on three cases including the bowtie-shaped apertures, the bowtie particles, and the full hybrid structure (DHS) consisting both. \textbf{Figure~\ref{aperturesim}a} shows the y- and x-polarization resolved simulation results for the apertures in the absence of bowtie particles. The simulations clearly show field (intensity) hotspots on the order of 50 and 100, for y- and x-polarized incident fields, respectively. As expected, the hotspots reside inside the aperture, which allows the signal to propagate to the collection optics (residing in the positive z-direction). One curiosity is the polarization dependence of the FE. Apparently the y-polarized incident light produces four high field intensity spots away from the bowtie center. Thus, these simulations confirm our first three hypotheses \textit{H1--H3}. Intriguingly, the complete hybrid structure in \textbf{Figure~\ref{aperturesim}b} produces approximately similar field enhancements at the aperture region, with the exception that the maximum x-polarized enhancement at the gap of the bowtie-shaped aperture is slightly higher (120 instead of 100), thus indicating that \textit{H4} might be valid as well. 

\begin{figure*}[t!]
    \centering
    \includegraphics[width=0.78\textwidth]{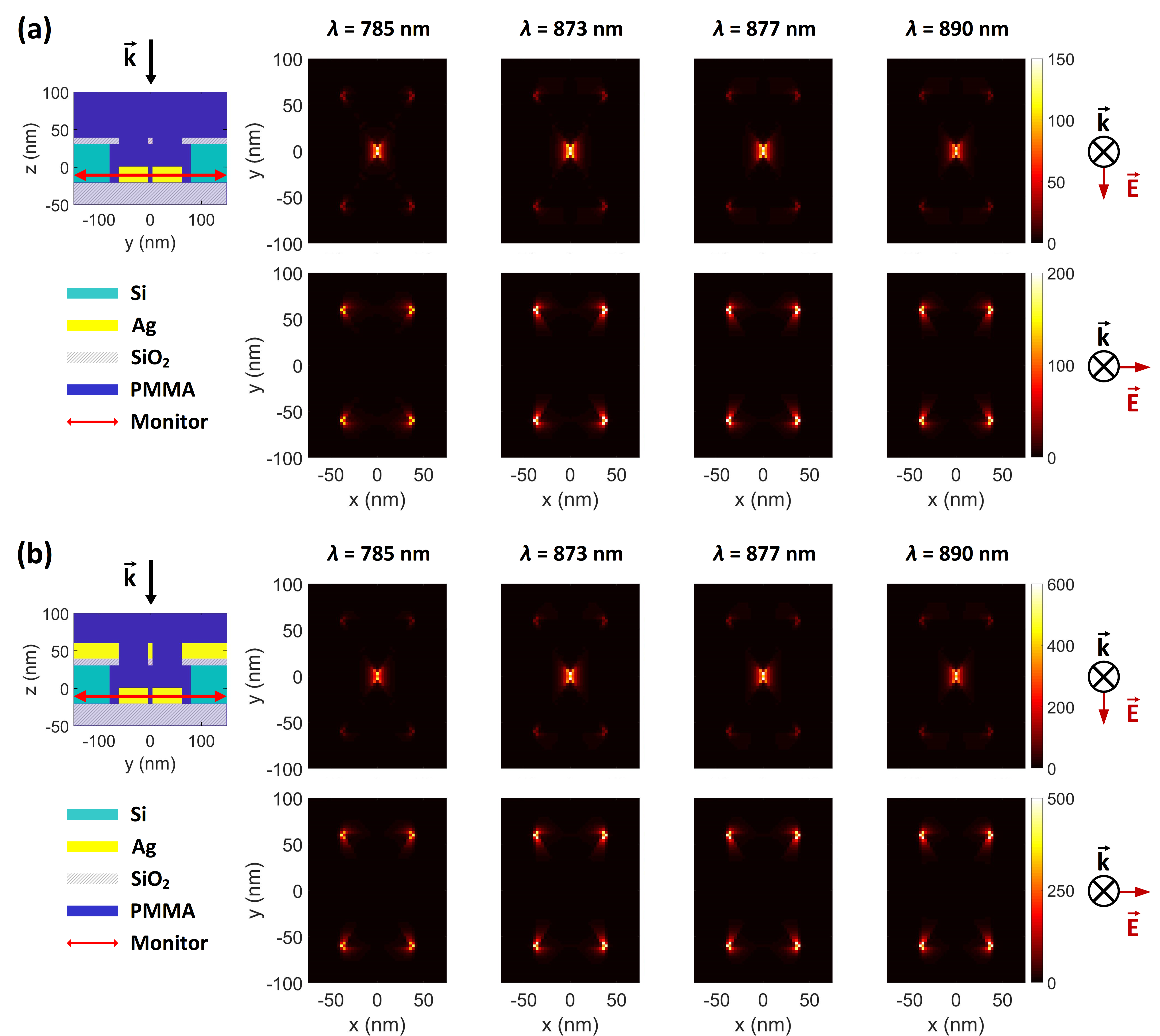}
    \caption{FDTD simulations showing electric field intensity profiles at the bowtie particles. The monitored area is indicated in the inset. (a) Bowtie only (i.e. DHS without the aperture) as a control. (b) DHS sample. The intensities are shown at the Raman excitation wavelength (785 nm) as well as at the Raman transition wavelengths of Rhodamine 6G (873 nm, 877 nm, and 890 nm) for both longitudinal (top panel) and transverse (bottom panel) polarizations.}
    \label{bowtiesim}
\end{figure*}

In \textbf{Figure~\ref{bowtiesim}} we compare the FEs associated with the bowtie particles and the hybrid structure. Our structure is designed such that the broad bowtie particle resonances overlap with the excitation light (785 nm) and the Raman transitions of R6G. In \textbf{Figure~\ref{bowtiesim}a} the sample containing only bowtie particles produces FEs on the order of 150--200 for both polarizations. Strikingly, the hybrid structure in \textbf{Figure~\ref{bowtiesim}b} exhibits enormous FEs of the order of 500--600 at the gap region of the bowtie particles. This indicates that the presence of the aperture layer in fact increases the FEs associated with the bowtie particles, fully confirming \textit{H4}. Further, we carried out simulations at an x-y plane residing between the bowtie particles and apertures as a control. These plots indicate that there exists a significant interlayer coupling between plasmon resonances of the bowtie particles and the apertures, see \textbf{Supporting Information Figure S1}. Notably, the sum of $\sqrt{\text{FE}}$ (which is equal to the electric field enhancement) for the bowtie-only and the aperture-only structures results in a smaller value than the $\sqrt{\text{FE}}$ of the hybrid structure. This suggests that the interlayer coupling could provide an additional enhancement for Raman signal measurements.

To experimentally evaluate the role of the aperture layer, we fabricated three sets of samples according to \textbf{Figure \ref{scheme}}. Our previously introduced BLIN method\cite{Piskunen2021Biotemplated} was used to make particle-aperture hybrid structures (BHS, \textbf{Figure \ref{scheme}b}) and plain bowtie particles (BBS, \textbf{Figure \ref{scheme}c}). Importantly, these two control sample sets allow a direct comparison between structures consisting of the bowtie particles only and the hybrid structures. The third set of samples was also comprised of hybrid structures, but they were fabricated \textit{via} a modified and optimized DALI process (DHS, \textbf{Figure \ref{scheme}a}). The advantage of DALI over the BLIN processing is the absence of thick PMMA and Si layers, which may then enable a stronger interlayer coupling between the aperture and the particles as shown in the simulations in \textbf{Figure~\ref{aperturesim}}, \textbf{Figure~\ref{bowtiesim}} and \textbf{Supporting Information Figure~S1}.

\begin{figure}[t!]
    \centering
    \includegraphics[width=\columnwidth]{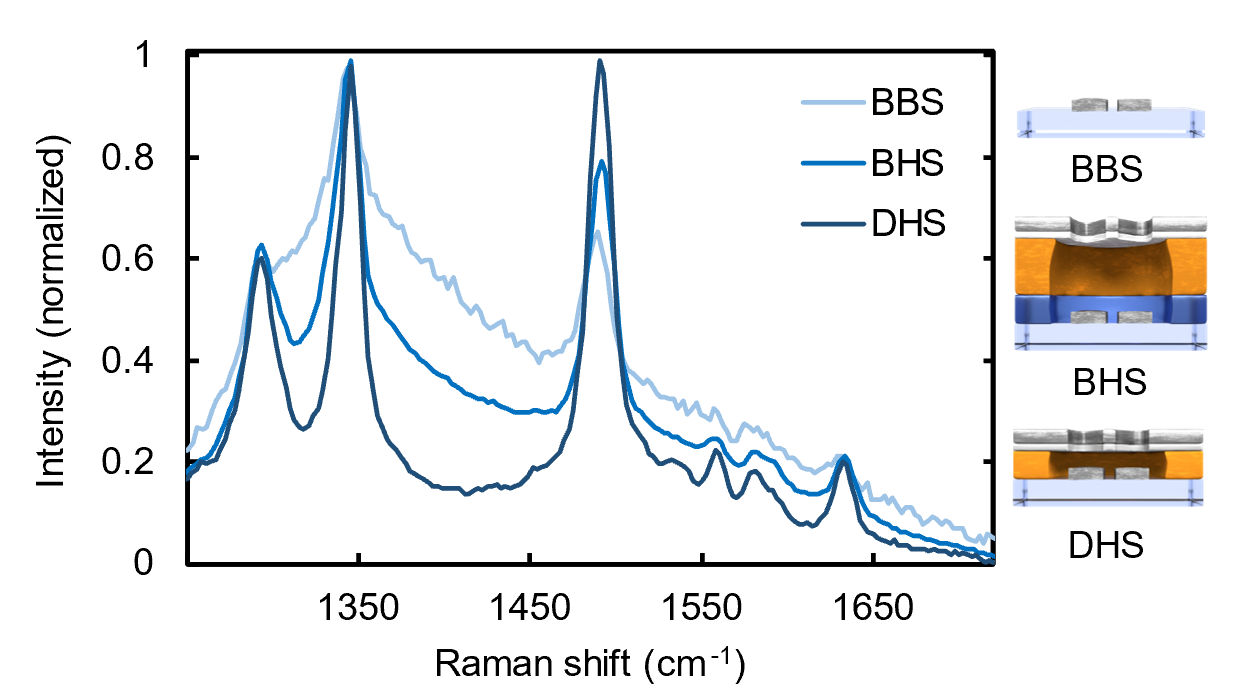}
    \caption{Comparison of normalized R6G Raman spectra measured on DHS, BHS and BBS substrates.}
    \label{ramanspectra}
\end{figure}

In \textbf{Figure \ref{ramanspectra}} we present the normalized Raman spectra for all three samples (BBS, BHS and DHS) overlaid with a layer of R6G-doped PMMA. From these three spectra we can distinguish very clear peaks at 1290, 1345, and 1490~cm$^{-1}$ (corresponding to the wavelengths of 873, 877, and 890 nm at the 785 nm excitation), which are associated with the prominent R6G Raman transitions.\cite{alam2020double} Due to practical reasons, we base our analysis here on normalized spectra, as the most relevant quantity, namely the signal-to-background ratio becomes most evident using this method. First, the BBS sample exhibits only very moderate Raman enhancements (light blue). Despite the significant background intensity, one can nevertheless distinguish the three relevant Raman peaks related to R6G. The presence of the silver layer, however, significantly improves the signal-to-background ratio as can be seen in the BHS sample (blue). We associate this with the significant background suppression (\textit{H1}) and additional hotspots related to the plasmonic enhancement in the aperture layer (\textit{H2,H3}).

Finally, the optimized DHS sample starts to reveal less pronounced background-obscured spectral features of R6G around 1600~cm$^{-1}$ (dark blue) as it takes the full advantage of all the four contributions of the new design, namely the suppressed background due to silver layer (\textit{H1}), selective aperture transmission from the plasmonic hotspots (\textit{H2}), plasmonic resonances related to nanoscopic apertures (\textit{H3}), and additional FE stemming from the enhanced coupling of the plasmonic resonances in the aperture and particle layer (\textit{H4}). This comparison of the spectra indeed manifests the outstanding performance of DHS compared to the control samples, especially to the commonly employed nanoparticle-based Raman substrates. Further, the optimized DALI structure clearly benefits from the increased interlayer coupling as compared to the BLIN reference sample.

\section*{Conclusions}

In summary, we have shown that the presented hybrid structuring of optically resonant substrates may have several benefits compared to conventional nanoparticle-based SERS surfaces. By using DALI-fabricated bowtie particles and bowtie-shaped apertures, we were able to achieve a four-fold advantage over particle-based substrates. In addition, we observed that the DALI-fabricated hybrid structures performed better than the BLIN-fabricated ones, indicating that the interlayer coupling strength between the particles and the apertures depends on the distance between the layers. This provides yet another degree of freedom in our system as it could allow tuning of the interlayer coupling by design. Currently, our method is based on single, discrete DNA origami structures, however, modular DNA origami units can also assemble into hierarchical arrays and macroscopic lattices.\cite{xin2021scaling} Therefore, the presented parallel and affordable\cite{praetorius2017biotechnological,zhang2023catalytic} DNA origami-based fabrication schemes could potentially be extended to highly ordered hybrid structures with even more intriguing optical features.

\section*{Supporting Information Available}

  Detailed lists of materials and equipment, process parameters, additional FDTD simulations (PDF)

\begin{acknowledgements}
The authors thank the Academy of Finland (project number 322002), the Emil Aaltonen Foundation, the Sigrid Jusélius Foundation, the Jane and Aatos Erkko Foundation, the Magnus Ehrnrooth Foundation, the Finnish Cultural Foundation (Kalle and Dagmar Välimaa Fund), and ERA Chair MATTER from the European Union’s Horizon 2020 research and innovation programme under grant agreement No 856705. The work was carried out under the Academy of Finland Centers of Excellence Programme (2022-2029) in Life-Inspired Hybrid Materials (LIBER), project number 346110. The work is part of the Academy of Finland Flagship Programme, Photonics Research and Innovation (PREIN), decision 321066. We also acknowledge the provision of facilities and technical support by Aalto University Bioeconomy Facilities, OtaNano - Nanomicroscopy Center (Aalto-NMC), and Micronova Nanofabrication Center.
\end{acknowledgements}

\begin{contributions}
Kabusure M. Kabusure: 0000-0003-4839-1189 \\
Petteri Piskunen: 0000-0002-3142-3191 \\
Jiaqi Yang: 0000-0001-9111-2769 \\
Veikko Linko: 0000-0003-2762-1555 \\
Tommi K. Hakala: 0000-0003-3853-4668  
\end{contributions}

\begin{interests}
The authors declare no competing financial interest.  
\end{interests}

\section*{Bibliography}
\bibliography{ms}

\end{document}


\title{\LARGE{Supporting Information} \\ \vspace{0.7 cm} Raman Enhancement in Bowtie-Shaped Aperture-Particle Hybrid Nanostructures Fabricated with DNA-Assisted Lithography}
\shorttitle{Supporting Information}

\author[1,$\dagger$]{Kabusure M. Kabusure}
\author[2,$\dagger$]{Petteri Piskunen}
\author[1]{Jiaqi Yang}
\author[2,3,4,*]{Veikko Linko}
\author[1,*]{Tommi K. Hakala}

\affil[1]{Department of Physics and Mathematics, University of Eastern Finland, Yliopistokatu 2, P.O Box 111, 80101, Joensuu, Finland}
\affil[2]{Biohybrid Materials, Department of Bioproducts and Biosystems, Aalto University, P.O. Box 16100, 00076 Aalto, Finland}
\affil[3]{LIBER Center of Excellence, Aalto University, P.O. Box 16100, 00076, Aalto, Finland}
\affil[4]{Institute of Technology, University of Tartu, Nooruse 1, 50411, Tartu, Estonia}
\affil[$\dagger$]{Equal contribution}
\affil[*]{Correspondence and requests for materials should
be addressed to \href{mailto:veikko.pentti.linko@ut.ee}{veikko.pentti.linko@ut.ee} or
\href{mailto:tommi.hakala@uef.fi}{tommi.hakala@uef.fi}}
\maketitle

\onecolumn

\newpage

\section{Materials and Equipment}

\vspace{0.5cm}

\begin{table}[h!]
\caption{Materials}
\label{Materials}
\centering
\bgroup
\def\arraystretch{1.5}%
\begin{tabularx}{\linewidth}{|c|c|c|X|}

\hline
\textbf{Name} & \textbf{Company} & \textbf{Catalog Number} & \textbf{Notes} \\ \hline
Acetone    & Honeywell    & 40289H    & Semiconductor grade ULSI, $\geq$99.5~$\%$    \\
Anisole    & MicroChem    &  MM030100   & A Thinner    \\
Ammonium hydroxide    & Fisher Chemical    & 10652251    & 25~$\%$, Certified AR for analysis, $d=0.91$      \\ 
Isopropanol    & Honeywell    & 40301H    & Semiconductor grade VLSI, $\geq$99.8~$\%$      \\
Menzel glass  & Thermo Scientific    &     & Borosilicate glass substrate, \newline 22 $\times$ 26~mm, 0.5--0.6 mm thickness      \\
Magnesium chloride    & Sigma Aldrich    & M8266    & Anhydrous, $\geq$99.8~$\%$      \\
Oligonucleotides    & Integrated DNA Technologies    &     & Staple strands for DNA origami bowtie, \newline 100~$\mu$M in RNase-free water    \\
p7249 ssDNA scaffold    & Tilibit Nanosystems    &     & Scaffold strand for DNA origami bowtie, 100 nM     \\
PMMA 950K in anisole   &   MicroChem  &     & A11 and A9 stock, diluted to target concentration with anisole      \\
Silver    & Kurt J. Lesker Company    & EVMAG40EXE    & 99.99~$\%$, PVD source      \\
Sodium chloride    & Sigma Aldrich    & S9888    & ACS reagent, $\geq$99.0~$\%$      \\
TAE buffer    & VWR Chemicals    & 444125D    & 50x concentrate at pH 8.0, \newline Electran Electrophoresis grade      \\
Tetraethyl orthosilicate    & Sigma Aldrich    & 86578    & $\geq$99.0~$\%$ (GC)      \\
Titanium    & Kurt J. Lesker Company    & EVMTI45EXE-A    & 99.995~$\%$, PVD source      \\ \hline
\end{tabularx}
\egroup
\end{table}

\vspace{0.5cm}

\begin{table}[h!]
\caption{Equipment}
\label{Equipment}
\centering
\bgroup
\def\arraystretch{1.5}%
\begin{tabularx}{\linewidth}{|c|c|c|}

\cline{1-3}
\textbf{Name} & \textbf{Company} & \textbf{Notes}  \\ \cline{1-3}
BRANSON 5510    & Branson        & Ultrasonic bath     \\
Dimension Icon    & Bruker        & Atomic force microscope   \\ 
Electron-beam evaporator IM-9912    & Instrumentti Mattila        & Tool for physical vapor deposition    \\ 
HBR 4    & IKA        & Heating bath   \\ 
Plasmalab 80+ PECVD    & Oxford Instruments        & Tool for plasma-enhanced chemical vapor deposition    \\ 
Plasmalab 80+ RIE    & Oxford Instruments        & Tool for reactive ion etching    \\
inVia Reflex    & Renishaw        & Raman microscope    \\ 
Sigma VP    & Zeiss        & Scanning electron microscope    \\ 

\cline{1-3}
\end{tabularx}
\egroup
\end{table}

\newpage
\section{Process parameters}

The film deposition and etching parameters were adapted from references [S1-S3] and adjusted when necessary for the current layer thicknesses. Ti and Ag films were deposited using slow evaporation rates ($\sim$0.1~Å/s)

\vspace{0.5cm}

{\centering
\begin{table}[h!]
\caption{Process parameters}
\label{Process parameters}

\RaggedRight{
\bgroup
\def\arraystretch{2}%
\begin{tabularx}{\linewidth}{|m{2.5 cm}|m{1 cm}|m{1.5 cm}|m{1.5 cm}|m{1 cm}|m{2 cm}|m{1.5 cm}|m{1.5 cm}|}

\cline{1-8}
 & \textbf{Gas} & \textbf{Gas flow \newline [sccm]} & \textbf{Chamber \newline pressure \newline [mTorr]} &\textbf{RF \newline power \newline [W]} & \textbf{Temperature \newline [\celsius]} & \textbf{Duration \newline [s]} & \textbf{Etch rate \newline [nm/min]} \\
\cline{1-8}
PECVD of a-Si (DHS) & SiH$_4$ \newline He & 14 \newline 266 & 2000 & 10 & 177 & 120 & - \\
\cline{1-8}
PECVD of a-Si (BHS/BBS) & SiH$_4$ \newline He & 14 \newline 266 & 2000 & 10 & 100 & 240 & - \\
\cline{1-8}
O$_2$ plasma & O$_2$ \newline Ar & 45 \newline 5 & 250 & 50 & - & 120 & - \\
\cline{1-8}
RIE of SiO$_2$ & CHF$_3$ \newline Ar & 25 \newline 25 & 40 & 30 & - & 35 & 10 \\
\cline{1-8}
RIE of a-Si \newline (DHS) & SF$_6$ \newline O$_2$ & 100 \newline 8 & 90 & 30 & - & 25 & 110 \\
\cline{1-8}
RIE of a-Si \newline (BHS/BBS) & SF$_6$ \newline O$_2$ & 100 \newline 8 & 90 & 30 & - & 60 & 110 \\
\cline{1-8}
RIE of PMMA & O$_2$ \newline Ar & 45 \newline 5 & 250 & 50 & - & 45 & 40 \\

\cline{1-8}
\end{tabularx}
\egroup
}
\end{table}\par
}

\newpage

\section{Additional FDTD Simulations}

\begin{figure}[h!]
    \centering
    \includegraphics[width=0.9\columnwidth]{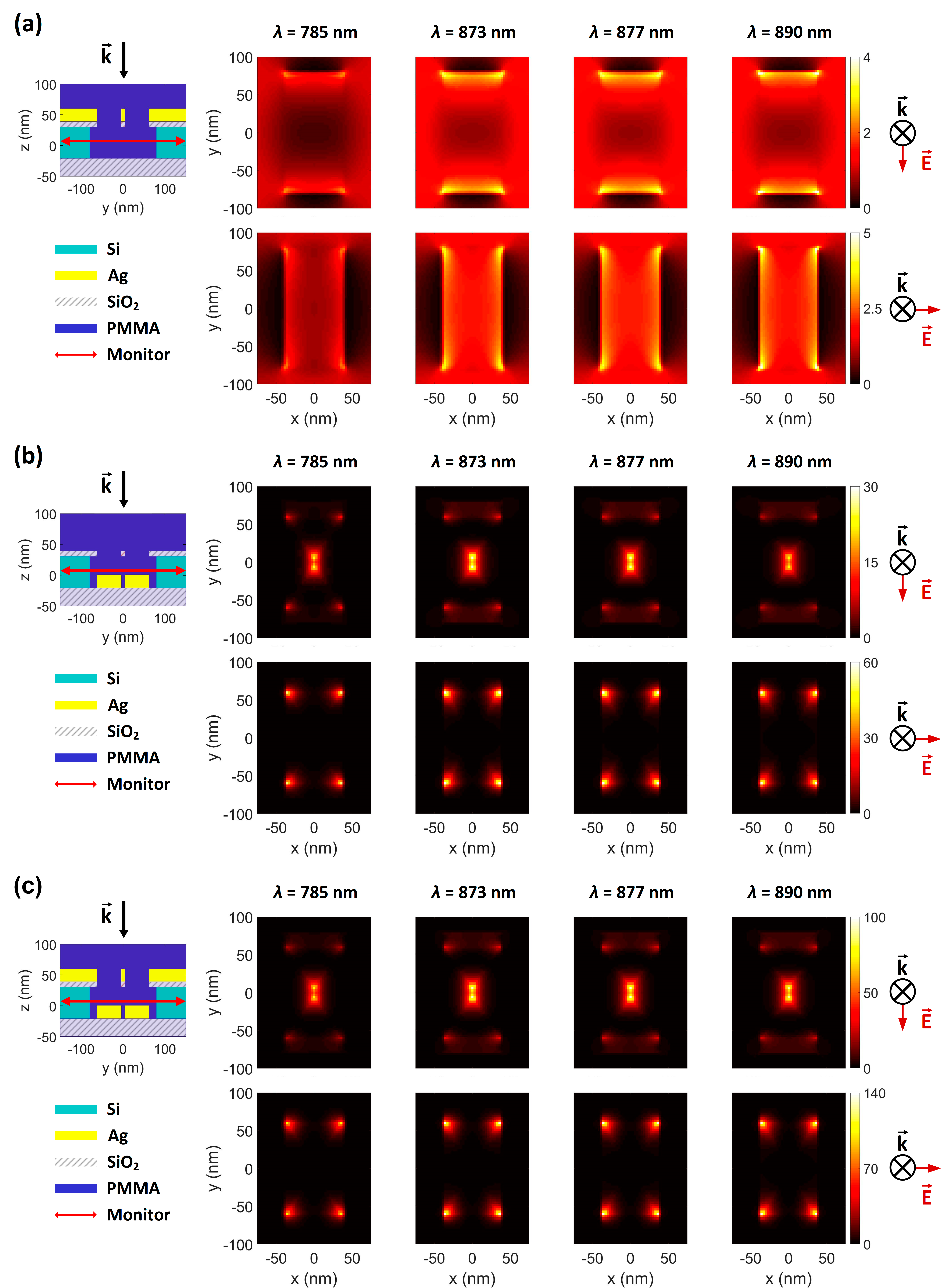}
    \caption{FDTD simulations showing electric field intensity profiles at the x-y plane residing between the bowtie particles and apertures. The monitored area is indicated in the inset. (a) Aperture only (i.e. DHS without the bowties). (b) Bowtie only (i.e. DHS without the apertures). (c) DHS sample. The intensities are shown at the Raman excitation wavelength (785 nm) as well as at the Raman transition wavelengths of Rhodamine 6G (873 nm, 877 nm, and 890 nm) for both longitudinal (top panel) and transverse (bottom panel) polarizations.}
    \label{TransEXsim}
\end{figure}

\newpage

\section{Supplementary Information References}

[S1] Shen, B.; Linko, V.; Tapio, K.; Pikker, S.; Lemma, T.; Gopinath, A.; Gothelf, K. V.; Kostiainen, M. A.; Toppari, J. J. Plasmonic Nanostructures through DNA-Assisted Lithography. \textit{Sci. Adv.} \textbf{2018}, \textit{4}, eaap8978.

\vspace{0.3cm}

[S2] Piskunen, P.; Shen, B.; Keller, A.; Toppari, J. J.; Kostiainen, M. A.; Linko, V. Biotemplated Lithography of Inorganic Nanostructures (BLIN) for Versatile Patterning of Functional Materials. \textit{ACS Appl. Nano Mater.} \textbf{2021}, \textit{4}, 529--538.

\vspace{0.3cm}

[S3] Kabusure, K. M.; Piskunen, P.; Yang, J.; Kataja, M.; Chacha, M.; Ojasalo, S.; Shen, B.; Hakala, T. K.; Linko, V. Optical Characterization of DNA Origami-Shaped Silver Nanoparticles Created through Biotemplated Lithography. \textit{Nanoscale} \textbf{2022}, \textit{14}, 9648--9654.